\documentclass[pre,aps,showpacs,showkeys]{revtex4}
\usepackage{amsmath}
\usepackage{amssymb}
\usepackage{graphicx}

\begin{document}
\title{Nonlinear Hamiltonian dynamics of Lagrangian transport and 
mixing in the ocean}
\author{M.V.~Budyansky, M.Yu.~Uleysky, S.V.~Prants}
\affiliation{Laboratory of Nonlinear Dynamical Systems,
V.I.Il'ichev Pacific Oceanological Institute of the Russian Academy of Sciences,
690041 Vladivostok, Russia}
\begin{abstract}
Methods of dynamical system's theory are used for numerical study of 
transport and mixing of passive particles (water masses, temperature, salinity, 
pollutants, etc.) in simple kinematic ocean models 
composed with the main Eulerian coherent structures in a randomly 
fluctuating  ocean~---  a jet-like current and an eddy.
Advection of passive tracers in a periodically-driven flow consisting of a 
background stream and an eddy (the model inspired by the phenomenon of 
topographic  eddies over mountains in the ocean and atmosphere) 
is analyzed as an example of chaotic particle's 
scattering and transport. A numerical analysis reveals
a nonattracting chaotic invariant set $\Lambda$
that determines  scattering and trapping of particles
from the incoming flow. It is shown that both the trapping time for particles
in the mixing region and the number of times their trajectories wind around
the vortex have hierarchical fractal structure as functions of the initial 
particle's coordinates. Scattering functions are
singular on a Cantor set of initial conditions, and this property should manifest
itself by strong fluctuations of quantities measured in experiments. 
The Lagrangian structures in our numerical
experiments are shown to be similar to those found in a recent laboratory
dye experiment at Woods Hole. Transport and mixing of passive 
particles is studied in the kinematic model inspired by the interaction 
of a jet current (like the Gulf Stream or the Kuroshio) with an eddy in 
a noisy environment. We demonstrate a 
non-trivial phenomenon of noise-induced clustering of passive 
particles and propose a method to find such clusters in numerical experiments. 
These clusters are patches of advected particles which can move together in a
random velocity field for comparatively long time. The clusters appear due
to existence of regions of stability in the phase space which is the physical 
space in the advection problem.
\end{abstract}
\keywords{Chaotic advection, coherent structures, noise-induced clusters}
\pacs{47.52.+j, 47.53.+n, 92.10.Ty}
\maketitle

\section{Introduction}
\begin{figure}[!htb]
\begin{center}
\includegraphics[bb=20 20 575 455,width=0.95\textwidth,clip]{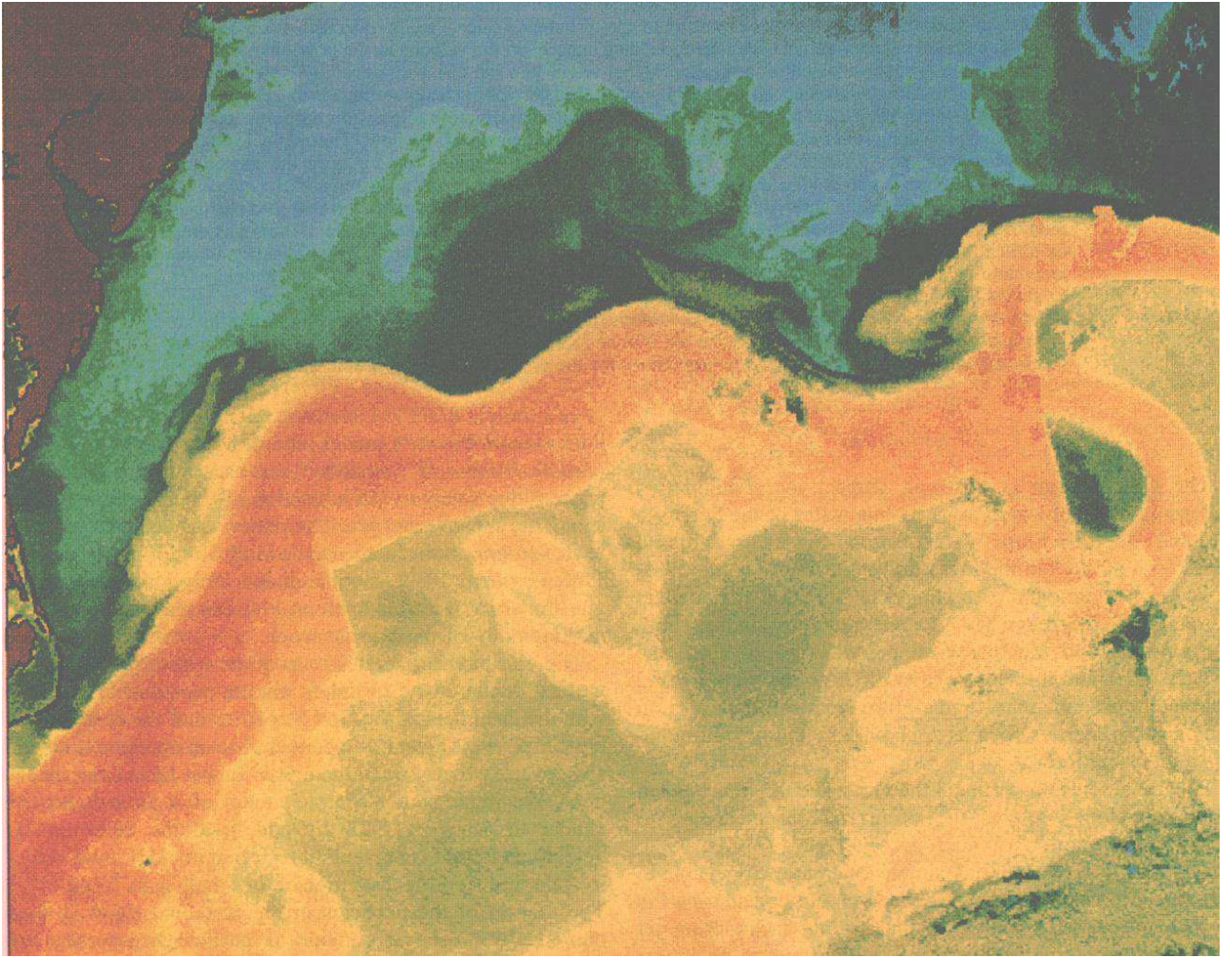}
\caption{Satellite image  of the Gulf Stream. Redder colors indicate 
warmer water. NOAA advanced very high resolution radiometer.}
\label{fig1}
\end{center}
\end{figure}

In the last decade, ideas and methods of dynamical systems theory have been used 
actively in physical oceanography with the aim to describe qualitatively and
quantitatively impact of coherent structures in the ocean on transport and
mixing of passive particles
(water masses, temperature, salinity, pollutants, etc.)
\cite{D00,B89,DW96,KK99,PD04,KJ04,KK04}. By passive particles, one means particles which take on the velocity
of a flow very rapidly and do not influence the flow. By a coherent structure, we
mean a quasistationary structure that can be recognized during a time much
longer than all  Eulerian time characteristics of a flow. Example of {\it an Eulerian
coherent structure} is shown in  Fig.~\ref{fig1} as a satellite image of the Gulf Stream
with a loop of warmer water. In the Lagrangian approach, we are interested not in the velocity
field but in movement of fluid parcels or passive scalars which satisfy the 
vector equation
\begin{equation}
\frac{d{\bf r}}{dt}={\bf v}({\bf r},t),
\label{1}
\end{equation}
where ${\bf r}(x,y,z)$ and ${\bf v}(u,v,w)$ are the vector position of a
particle and its vector velocity at the point $(x,y,z)$, respectively. 
The Eulerian
velocity field $\bf v$ is given as a solution of a hydrodynamic equation or
it is measured in the ocean (on the sea surface it can be done, for example, 
with the help of a Doppler radar). 
Eq.~\ref{1} is a three dimensional dynamical system whose phase
space is the physical space for advected particles. It is well known in dynamical
system's theory that invariant manifolds, like stationary points, different kinds
of attractors, including strange ones, KAM tori, cantori, stable and unstable
manifolds, define mainly transport and mixing in the phase space. In hydrodynamics
it is natural to call them {\it Lagrangian coherent structures} that define global
transport and mixing of passive particles in fluids. Lagrangian structures can be
visualized in laboratory in dye experiments and partially in the ocean with the
help of drifters and buoys \cite{D91,RPW86}.
The main task of dynamical system's theory in physical oceanography is to
find and specify Lagrangian coherent structures and their impact on transport
and mixing of water masses. Methods of that theory are attractive especially
because they are common for different classes of models, dynamic and kinematic
ones, and do not depend on their specific analytic forms because they use,
mainly, fundamental geometric structures.

In this paper in the framework of the geometrical approach, we describe transport
and mixing of passive particles in  geophysical flows composed with the main Eulerian
coherent structures in the ocean, a  jet-like current and an
eddy. The focus of this work is not on dynamics of the flow but rather on
mechanisms of particle's transport and mixing. That is why we use, as a prototype 
model, a kinematic
model that was introduced in \cite{JTPL01} to describe the phenomenon of
topographical eddies in the ocean \cite{K83,Z95,H73,GC97}. A current 
over seamounts and  bottom cavities produce on the rotating Earth  
(quasi)stationary 
anticyclonic and cyclonic eddies, respectively. Such eddies may arise over 
underwater ridges and valleys as well. 

In modeling oceanic flows we should take into account, generally speaking, 
 a periodic tidal
component of the current under consideration and a random component caused
by turbulent diffusion. A simple stream function that is able to describe
the main features of the two-dimensional flow of ideal fluid over a 
$\delta$-like bottom topography has the following normalized form:
\begin{equation}
\displaystyle{
\Psi=\ln{\sqrt{x^2+y^2}}+\varepsilon x+
\xi x\left[(1-\alpha)\sin\tau+\alpha F(\tau)\right]},
\label{2}
\end{equation}
where the first term represents a fixed point vortex placed at the point
with Cartesian coordinates $x=y=0$, the second and third ones  describe  
a steady and  periodic components of the current with the normalized velocities 
$\varepsilon$ and $\xi$, 
respectively, and $F(\tau)$ is a random function with $\alpha$ being
a normalized strength of noise. 

\section{Geometry of transport and mixing in a periodically-driven flow: 
jet current with a tidal component over delta-like bottom topography}  

It immediately follows from incompressibility and two-dimensionality of the
flow that equations of advection of passive particles are Hamiltonian ones
\begin{equation}
\begin{aligned}
\dot x&=u(x,y,\,\tau)=-\frac{\partial \Psi}{\partial y},\\
\dot y&=v(x,y,\,\tau)=\frac{\partial \Psi}{\partial x},
\end{aligned}
\label{3}
\end{equation}
where dot denotes differention with respect to the normalized time $\tau$. 
Analysis
of transport and mixing of advected particles in the flow given by the stream
function (\ref{2}) {\em without random force} has been done in our papers 
\cite{PD04,JETP04}. In this section we discuss briefly the main results in
the case of a periodic perturbation.
\begin{figure}[!htb]
\begin{center}
\includegraphics[width=0.8\textwidth,clip]{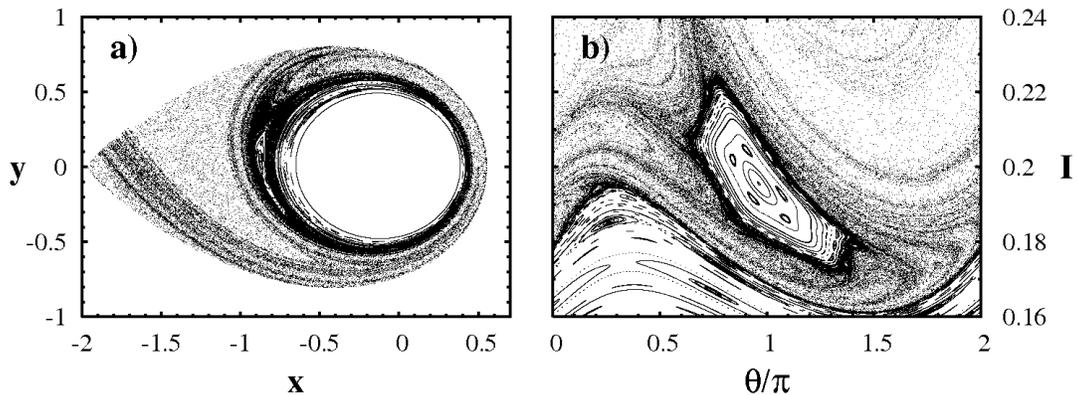}
\caption{Poincar\'e sections of the mixing zone in the Cartesian (a) and 
the action-angle (b) coordinates.}
\label{fig2}
\end{center}
\end{figure}
In the absence of any time dependence in the streamfunction (\ref{2}), the phase
portrait of the unperturbed system consists of finite and infinite orbits
separated by a separatrix encompassing the vortex and passing through the saddle 
point ($x=-1/\varepsilon$, $y=0$). In the polar coordinates
($x=\rho\cos\varphi$, $y=\rho\sin\varphi$),
the unperturbed equations are solved in quadratures
\begin{equation}
\dot\rho=\pm\varepsilon\,
\sqrt{1-\left(\frac{E-\ln \rho}{\varepsilon \rho}\right)^2},
\label{4}
\end{equation}
where $E=\varepsilon\rho\cos\varphi+\ln\rho$
is an integral of motion.
Depending on initial conditions, particles move either around the vortex
along closed streamlines encompassed by the separatrix loop or around
the loop along infinite streamlines. We have shown analytically and numerically
in \cite{JTPL01} that under a small perturbation there exist transversal
intersections of stable and unstable saddle-point manifolds in the neighborhood
of the unperturbed separatrix. Under periodic perturbation, the particle's 
trajectories deviate from the steady-flow streamlines. We define there zones
in the phase (physical) space as follows: free-stream region (with incoming and outcoming
components), mixing region and vortex core as the sets of trajectories with
the number of times they wind around the vortex being zero, finite and infinite,
respectively. Transport  is trivial in the free-stream region. The size of the 
vortex core depends on the ratio $\varepsilon/\xi$. The vortex core consists of regular
periodic and quasiperiodic trajectories with thin stochastic layers between
them. As the value $\varepsilon/\xi$ increases, the vortex core grows while the 
mixing zone
shrinks, correspondingly. The frequency of particle's rotation in the vortex 
core depends on the distance from the singular point $(0,0)$ and is much higher
than the perturbation frequency that is equal to $1$. Therefore, the 
perturbation can be treated as adiabatic with respect to most orbits inside
the core, and these orbits are regular, except for those lying in a  
neighborhood
of higher-order overlapping resonances which make up very narrow stochastic
layers. In applications, it is important that KAM tori in the vortex core 
make up impermeable barriers that limit particle's transport and mixing.

The topology of trajectories in the mixing zone is much more complicated.
In Fig.~\ref{fig2}  we show the Poincar\'e sections of the whole mixing region 
(Fig.~\ref{fig2}a  
with Cartesian coordinates $x$ and $y$) and of a part of this region in the 
neighborhood of the half-integer primary resonance (Fig.~\ref{fig2}b with the action
$I$ and angle $\vartheta/\pi$ variables) which is surrounded by higher order 
resonances.
High density of points around the vortex core and the islands means the presence
of cantori there which can trap particles for a long time. The cantori should
play an important role in transport and mixing of passive particles in 
geophysical flows. 
\begin{figure}[!htb]
\begin{center}
\includegraphics[width=0.8\textwidth,clip]{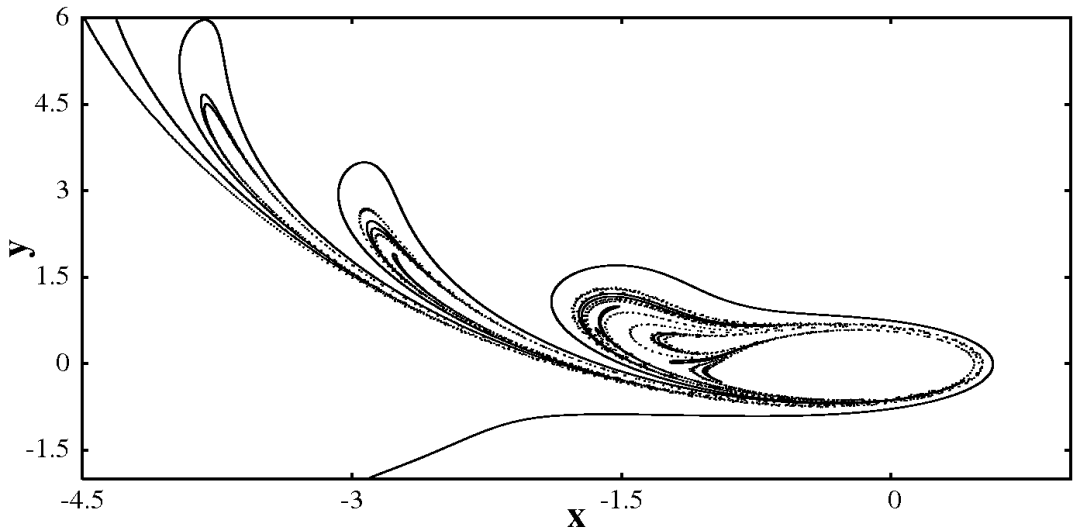}
\caption{Image of the unstable manifold $\Lambda_u$ obtained as a snapshot 
of a dye streak at $\tau=15\pi$.}
\label{fig3}
\end{center}
\end{figure}

A chaotic invariant set $\Lambda$ is defined as a set of all orbits
(except for the KAM tori and cantori) that never leave the mixing region.
The set consists of an infinite number of periodic and aperiodic 
(chaotic) orbits. All orbits in this set are unstable. If a tracer belongs
to $\Lambda$ at the initial moment, then it remains in the mixing region
as $\tau \to \infty$ or $\tau \to -\infty$. The Poincar{\'e} section of
$\Lambda$ is a set of points of Lebesgue measure zero. Most trajectories
of the tracers advected in the mixing region of the incoming flow sooner or later 
leave the mixing region with the outgoing flow. However, their behavior is
largely determined by the presence of $\Lambda$. They can ``trail'' after
trajectories of the saddle set wandering in their neighborhoods.

Each orbit in the chaotic set and, therefore, the entire set $\Lambda$
have both stable and unstable manifolds. The stable manifold
$\Lambda_s$ of the chaotic set is defined as the invariant set of
orbits approaching those in $\Lambda$ as $\tau \to \infty$.
The unstable manifold $\Lambda_u$ is defined as the stable 
manifold corresponding to time-reserved dynamics.
Following trajectories in $\Lambda_s$, tracers, advected by
the incoming flow, enter the mixing region and remain there forever.
It was mentioned above that the corresponding initial conditions
make up a set of measure zero. The tracer trajectories that are
initially close to those in the chaotic set follow the chaotic-set
trajectories for a long time and eventually deviate from them,
and leave the mixing region along the unstable manifold.
This behavior offers a unique opportunity to extract
important properties of $\Lambda$ by measuring the characteristics
of scattered particles and to observe unstable manifolds directly in 
laboratory experiments and even in geophysical flows.

An unstable manifold can be visualized by various methods.
A blob consisting of many tracer particles, initially belonging
to the intersection of the incoming flow with the stable 
manifold, spreads out and transforms into an intricate fractal
curve approaching $\Lambda_u$ in the course of time. A similar
pattern develops in dyeing experiments. The stable manifold 
lies in the coordinate-plane region bounded by the separatrix 
locations at the times corresponding to the two extrema 
reached during the perturbation period. This region extends
to $-\infty$ along the $y$ axis, and its width is determined by the
values of $\varepsilon$ and $\xi$. Only particles located in this 
region reach the mixing region. Figure ~\ref{fig3}  shows an image of the
unstable manifold at $\tau=15\pi$ obtained numerically
by integrating the equations of motions for particles continuously
injected into the incoming flow at the point with $x_0=-4.357759744$
and $y_0=-6$. This pattern oscillates with the period of the flow. Tracer particles 
are advected along the fractal curve of the unstable manifold,
which plays the role of an ``attractor'' in a Hamiltonian system \cite{JETP04} 
(there are no ``classical'' attractors in an incompressible flow).
\begin{figure}[!htb]
\begin{center}
\includegraphics[width=0.8\textwidth,clip]{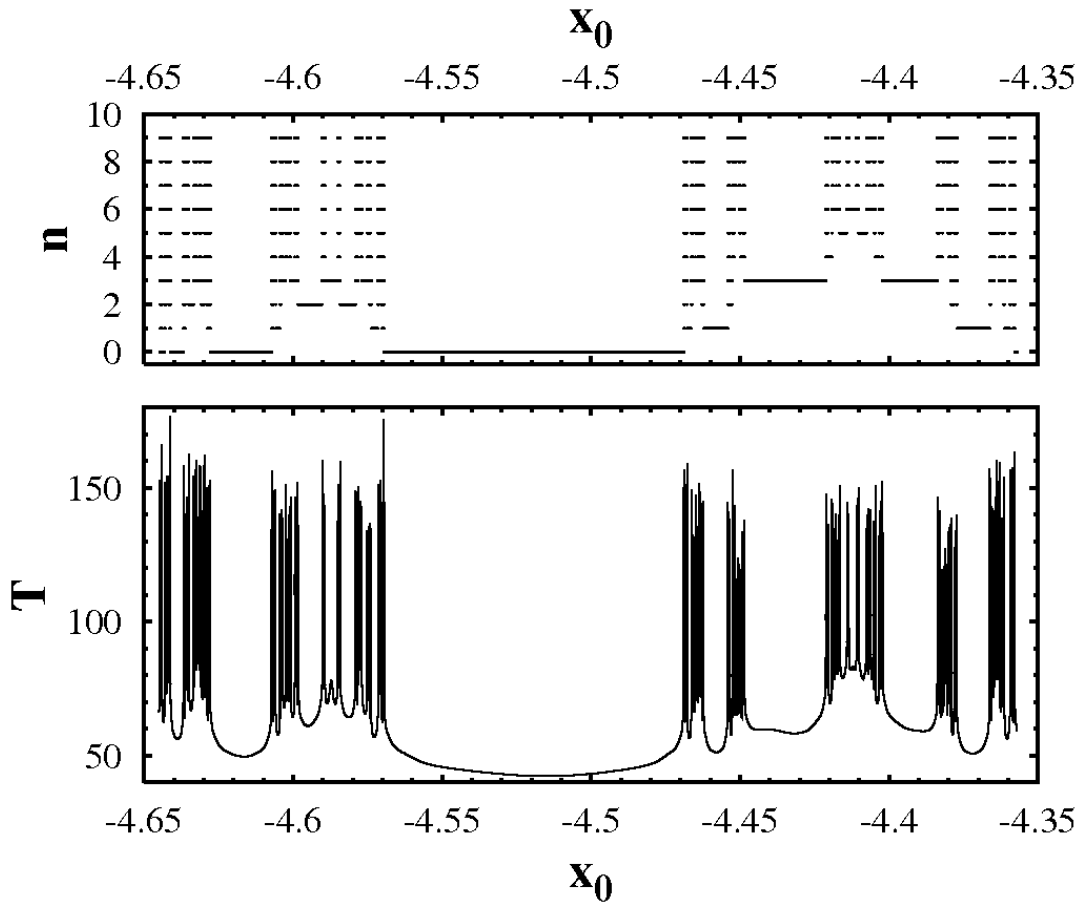}
\caption{Advection fractal under a periodic perturbation.
(a) the number of turns of particles around the 
vortex $n$ and (b) the respective exit time $T$  vs initial particle's 
position $x_0$.}
\label{fig4}
\end{center}
\end{figure}

Choosing a material line in the incoming flow,   
we calculate the total number $n$ of turns 
executed by 
particles, belonging to this line, before they escape into the 
outgoing-flow region. The graph of
$n(x_0)$ is shown to be an intricate self-similar hierarchy of sequences of
fragments of the material line (Fig.~\ref{fig4}a). Their fractal properties
are generated by the infinite sequence of intersections of stable and 
unstable manifolds with the material line segment as it 
rotates about the vortex.
Following to the paper \cite{M03}, we refereed to the sequences of segments
corresponding to each $n\ge 0$ as epistrophes. The epistrophes 
make up a hierarchy. 
Numerical experiments on epistrophes lying on
different levels reveal  the following trends:
(I) each epistrophe converges to a limit point in the material 
line segment under consideration; (II) the end points
of each segment in an $n$th-level epistrophe are the limit
points of an $(n+1)$th-level epistrophe; (III) the lengths of
segments in an epistrophe decrease in geometric progression; 
(IV) the common ratio $q$ of all progressions is related 
to the largest Lyapunov exponent for the saddle
point as follows: $\lambda=-(1/2\pi)\ln q$. The epistrophic structure 
manifests itself in the plot of dependence of the time of exit of 
particles from the mixing zone $T$ on their initial positions in 
the material line $x_0$. This plot, shown in Fig.~\ref{fig4}b,
demonstrates 
wild oscillations of the exit time $T$ in the neighborhood of 
singular points of the Cantor-like set which, in principle, 
can be measured in real laboratory experiments.  

Transport of passive particles is defined by a generic
mechanism of scattering and folding in the flow, where stretching is caused
by the stationary current component and folding~---
by the nonstationary tidal component and vortex. 
Strictly speaking, the structure of the fractal chaotic scattering
is defined by the chaotic invariant set $\Lambda$
with a stable $\Lambda_s$ and an unstable $\Lambda_u$ manifolds.
In the physical space $\Lambda_s$ acts as a variety of dynamical traps
that are able to trap the particles for a while. We have identified
a mechanism of trapping of the ends of strophes and epistrophes segments
with the saddle unstable periodic orbit. Stable manifolds
of the other unstable periodic and chaotic orbits ``attract''
advected particles and deposit in the structure of the fractal
as well. Effect of different dynamical traps manifests 
itself on different timescales and for different values
of the winding number $n$. For $\tau$ of the order of a few perturbation 
periods and small $n$, the mechanism of stretching and folding
defines the particle's transport. With increasing $\tau$ and $n$,
the saddle orbit and other unstable periodic and chaotic orbits, belonging
to the chaotic invariant set $\Lambda$, begin to play a significant
role in forming the fractal. 

\subsection{Transport and mixing of tracers in a Woods Hole laboratory experiment}

In this section we discuss the results of a laboratory experiment on chaotic
advection which has been carried out at the Woods Hole oceanographic Institution
\cite{DPH02} with the aim to find a possible mechanism for fluid transport and
mixing among a western boundary current and subbasin recirculation gyres. Possible
applications include the North Atlantic Deep Western Boundary Current and its 
adjacent mesoscale recirculation gyres. Transport and mixing of a dye in this
real experiment are shown to be very similar to those in our model flow 
described in the preceding section. The
respective flow has been set up in cylindrical laboratory tank (42.5 cm in
diameter and a mean depth of 20 cm) with a slopping bottom (a slope is $s=0.15$).
The tank rotates at a fixed rate $\Omega=2$ rad$\cdot$s${}^{-1}$ and the circulation
is driven by a lid that
rotates at a differential rate $\Delta\Omega<0$, producing a uniform, 
anticyclonic surface
stress vortex. Under a steady lid forcing, the (nearly) steady flow, consisting
of the western boundary current with the velocity of $u_\omega\simeq 0.1\div0.3$
cm$\cdot$s${}^{-1}$
and the twin gyre, can be 
set up. The main control parameter in the experiment is the ratio between the 
Stommel and inertial boundary layers $\delta\simeq 8\sqrt{\Delta\Omega/\Omega}$.
\begin{figure}[!htb]
\begin{center}
\includegraphics[width=0.8\textwidth,clip]{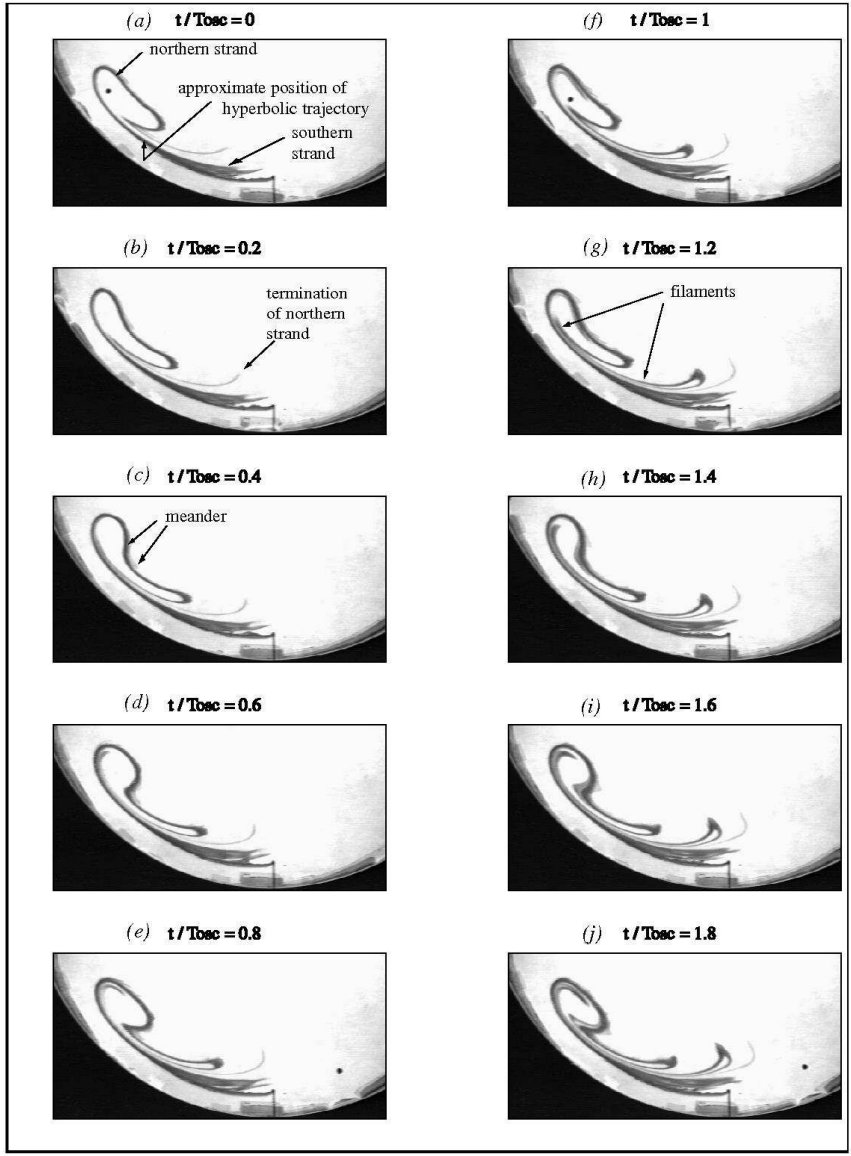}
\caption{Evolution of dye lines over two forcing cycles in the 
Woods Hole laboratory experiment \cite{DPH02}.} 
\label{fig5}
\end{center}
\end{figure}
The flow is almost steady if $\delta\leq 1.1$. Varying the lid rotation
\begin{equation}
\Delta(t)=\Delta\Omega_0\left(1+A_\text{osc}\sin\frac{2\pi t}{T_\text{osc}} \right),
\label{8}
\end{equation}
it is possible to produce a time-periodic flow. Different methods have been
used for visualizing and measuring the horizontal circulation. Neutrally buoyant
plastic particles are suspended in the tank and illuminated by a laser light.
Streak images showing the trajectories of the particles over short time intervals
are made from digital video recording of the flow. Direct velocity measurements
are made using particle image velocimetry with a spatial resolution of 1 cm. 
Another visualization technique involves the injection of dye into the flow
using a needle. Dye patterns are illuminated from below and recorded from above using
a video camera.

Within the range $1.1<\delta< 1.4$, the recirculation splits into twin gyres, the northern and
southern ones, forming the $\infty$-like figure. Mixing of the dye is weak 
in the steady
regime and thought to be due to double diffusion between the salty ambient water
and the food color dye. The time dependence introduces considerably more complicated
mixing of the dye. The sequence in Fig.~\ref{fig5}  \cite{DPH02} shows the evolution of the
dye contour over two periods of the lid oscillation with $\delta=1.25$, $A_\text{osc}=0.05$
and $T_\text{osc}=131$ s.
Dye is injected very close to the southern loop of unperturbed $\infty$-like separatrix
(south is to the right and west is downward in the figure). Fig.~\ref{fig5}a  corresponds
to the time moment about three cycles after injection is begun ($t/T_\text{osc}=0$),
and Fig.~\ref{fig5}j is
for $t/T_\text{osc}=1.8$. In fact, the above pictures illustrate the evolution 
of the unstable
manifold of the respective chaotic invariant set. 
\begin{figure}[!htb]
\begin{center}
\includegraphics[width=0.67\textwidth,clip]{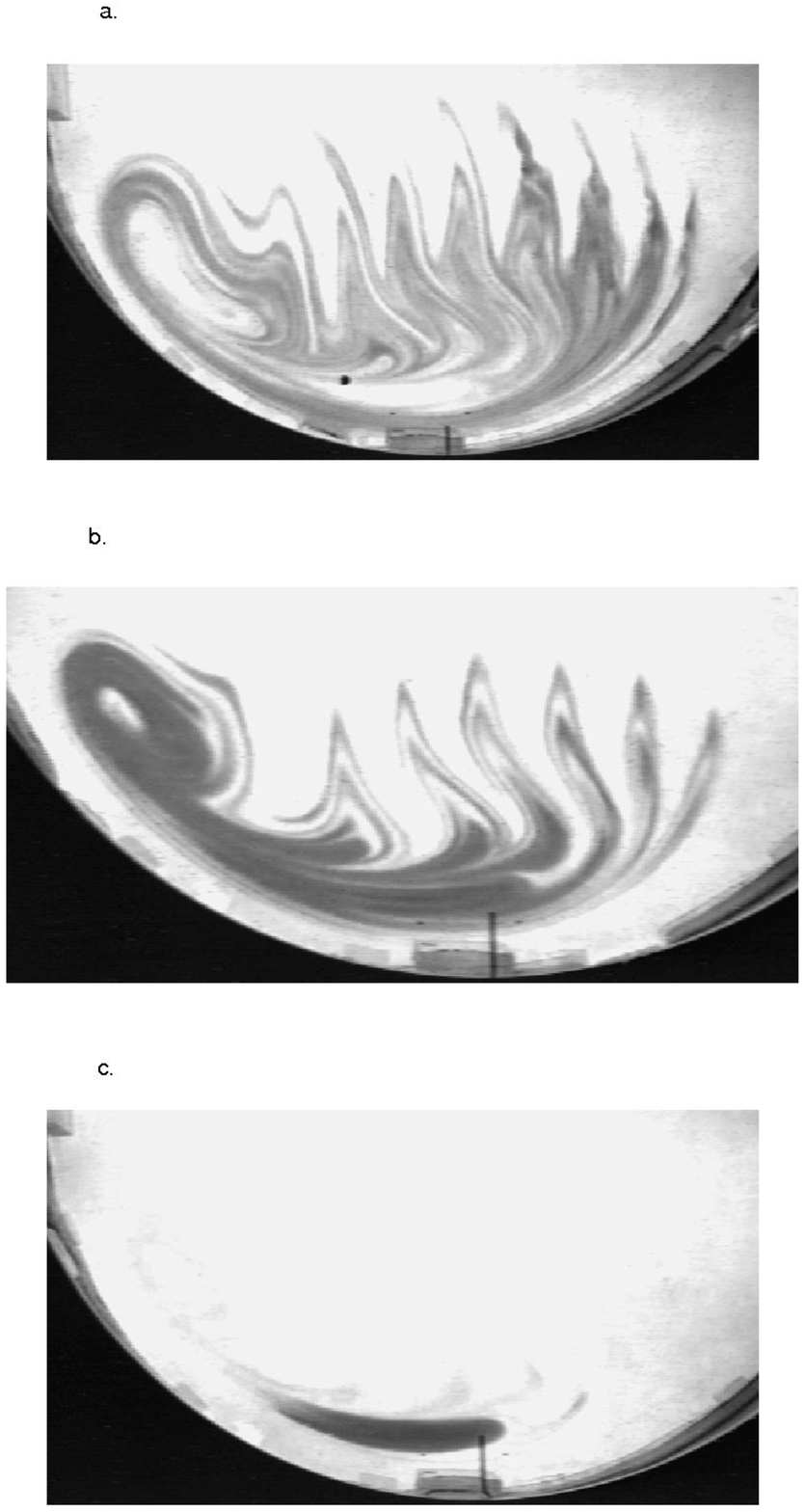}
\caption{Dye patterns with different places of injection in the 
Woods Hole laboratory experiment \cite{DPH02}.} 
\label{fig6}
\end{center}
\end{figure}
Transport and mixing depend strongly on which place in the flow the dye is
injected. Fig.~\ref{fig6}  \cite{DPH02} demonstrates the tracer streaklines with three
different places of injection: just outside the southern gyre (Fig.~\ref{fig6}a), 
just
inside the western edge of the southern gyre (Fig.~\ref{fig6}b) and inside the very
center of the southern gyre (Fig.~\ref{fig6}c). In the first two cases, the prominent
filaments of the dye are seen both in the southern and northern gyres. That is 
because the dye is injected very close to the unperturbed separatrix in both
the case. In the third case, the dye is injected inside the gyre core and
remains confined there by KAM-tori which are impenetrable barriers for
mixing (except for molecular diffusion).

We would like to pay attention on a similarity between the streaklines in
our numerical experiment (Fig.~\ref{fig3}) and the dye streaklines in the laboratory
experiment (Fig.~\ref{fig6}) \cite{DPH02}. In both the experiments, the dye is injected
nearly the unperturbed separatrix, and both the pictures are images of the
unstable manifolds of the respective chaotic invariant set $\Lambda$. 
A difference is that
the unperturbed separatrix in our model flow is of a $\propto$-like form, 
and, therefore,
our flow has only one gyre. The detailed analysis, of transport and mixing,
similar to one that has been done in our papers \cite{PD04,JETP04} and in this
paper, can be done for the laboratory flow \cite{DPH02} modeling transport and
mixing of water masses among western boundary currents and subbasin recirculation
gyres. Therefore, we expect for all the features of chaotic scattering:
fractals, anomalous transport, stickiness, etc.  

\section{Transport and mixing in a kinematic model of the current-eddy 
interaction in a noisy environment}
\subsection{Modeling noise}
       
Real flows, of course, have noisy components due to environment.
We will model noise $F(\tau)$ as a function consisting of
a thousand of harmonics 
\begin{equation}
\displaystyle{
F(\tau)=\frac{1}{\sqrt{N+1}}\sum\limits_{k=0}^{N}\sin(\omega_k\tau+\varphi_k)},
\label{F}
\end{equation}
with random phases $\varphi_k$ equally distributed in the range $[0:\,2\pi]$ 
and with frequencies $\omega_k=
\omega_b+k(\omega_e-\omega_b)/N$ equally distributed in a wide range  between
the lowest $(\omega_b)$ and highest $(\omega_e)$ frequencies.
It is possible, in principle, to model the spectrum of an arbitrary
form with the help of the series (\ref{F}) introducing an amplitude
factor depending on the frequency. Accordingly to the central limit
theorem, $F(\tau)$ has a Gaussian distribution with zero expectation value
because it is a large number of independent random variables.
The factor $1/\sqrt{N+1}$ provides the variance to be equal
to $1/2$.
\begin{figure}[!htb]
\begin{center}
\includegraphics[width=0.8\textwidth,clip]{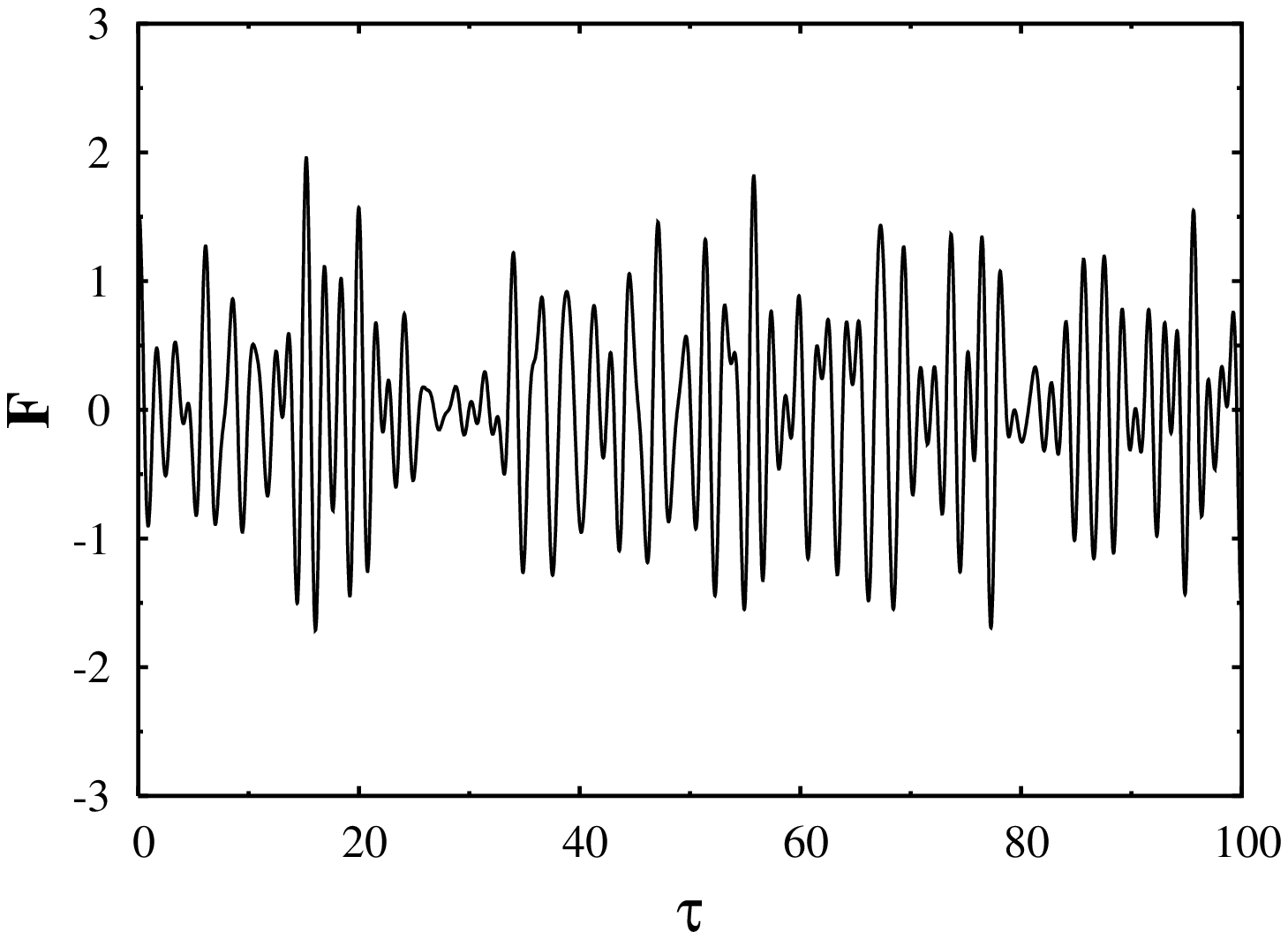}
\caption{A realization of the random function $F(\tau)$ with 
$\omega_b=2$, $\omega_e=5$ and $N=1000$.}
\label{fig7a}
\end{center}
\end{figure}
\begin{figure}[!htb]
\begin{center}
\includegraphics[width=0.8\textwidth,clip]{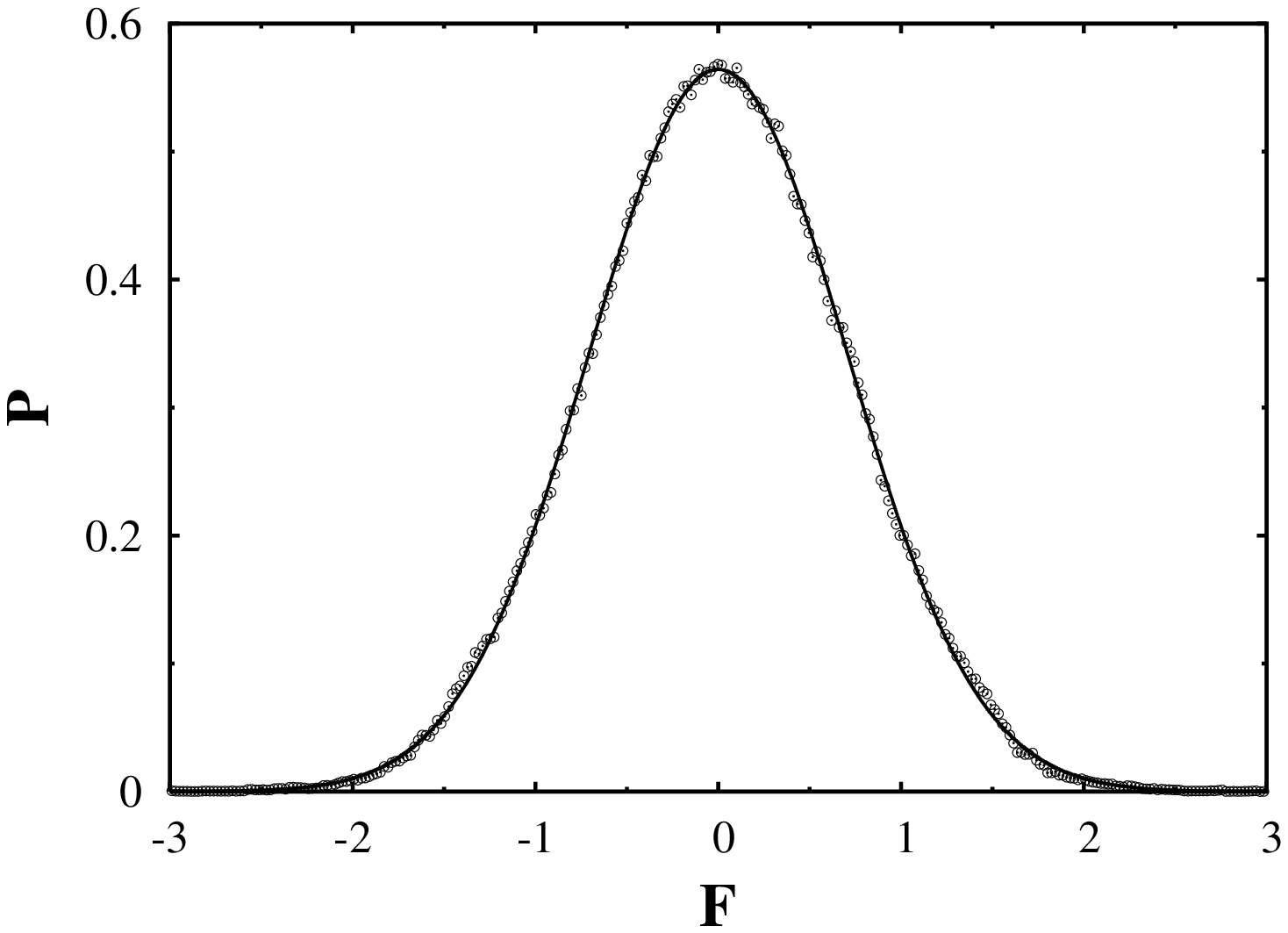}
\caption{PDF for the realization shown in Fig.~\ref{fig7a} (circles). 
Gaussian distribution with $\sigma^2=1/2$ is shown by the solid line.}
\label{fig7b}
\end{center}
\end{figure}

In Fig.~\ref{fig7a} we show a piece of a realization of the function 
$F(\tau)$ and in Fig.~\ref{fig7b} --- a probability distribution for this 
realization. In simulation we use not the function $F(\tau)$ itself  
but its approximation by cubic splines providing the computation time to be 
almost independent on the number of harmonics of $F(\tau)$.

Modeling noise with the help of the series (\ref{F}), instead of 
solving respective stochastic equations, has both the advantages and  
disadvantages. Among the advantages we mean the following: 
\begin{itemize}
\item The function  $F(\tau)$ is smooth. It allows to use in modeling 
equations with noise all the numerical methods and codes tested with 
problems without noise. 
\item Reproducibility. The function  $F(\tau)$ depends on a set of phases 
$\varphi_k$ which is produced by a random-number generator. Initializing the 
generator with the same number, we reproduce the same realization of 
$F(\tau)$ under the same $\omega_b$, $\omega_e$ and $N$. In such a way we 
can do different numerical experiments with the same realization of noise. 
\item Independence of solutions on the integration step. In difference from  
numerical solving of stochastic equations with additional dependence on 
the integration step, our method does not contain any additional dependence on 
the integration step. 
\item Noise with an arbitrary spectrum can be modeled.
\item A nonzero correlation length can be chosen to model noise with 
specific correlation properties. 
\end{itemize}
The following disadvantages of  our modeling of noise should be mentioned:
\begin{itemize}
\item It is necessary to cut the respective spectrum from 
above and below. 
\item Computing a high-frequency noisy perturbation requires 
more spline coefficients to memorize and more short integration steps. 
\item Spectrum of noise is not smooth but contains  a large number of 
delta-like peaks.  
\end{itemize}

\subsection{Noise-induced transport and mixing}

In this subsection we study  transport and mixing of passive particles 
in a kinematic model inspired by the interaction of a jet current 
(like the Gulf Stream or the Kuroshio) with an eddy (see Fig.~\ref{fig1}). 
Oversimplifying the problem, we model the respective flow by 
the stream function with a noisy component induced in the ocean by 
turbulence with different scales   
\begin{equation}
\displaystyle{\Psi_{noise}=\ln{\sqrt{x^2+y^2}}+\varepsilon x+
\xi x F(\tau)}.
\label{N}
\end{equation}
What is the effect of a broad-band weak noise on transport and 
mixing of passive particles?
What happens with the main structures and properties of the flow --- the 
vortex core, chaotic invariant set and its stable and unstable manifolds, 
fractals, anomalous transport, and main resonances --- on reasonable
timescales under a noisy-like excitation? 

The most evident effect of noise is fussing of smooth invariant curves of 
a deterministic Hamiltonian system under consideration. In the limit 
$\tau \to \infty$, no regions in the phase space remain forbidden.  
Even a weak noise can induce transport of particles through KAM~tori 
which cease to be impenetrable barriers to transport. Noise-induced diffusion 
depends mainly on the frequency range of noise and occurs 
on different time scales. For example, a vortex core may remain an 
impenetrable barrier for a long time comparing with a period of rotation 
of particles around the vortex. 
\begin{figure}[!htb]
\begin{center}
\includegraphics[width=0.8\textwidth,clip]{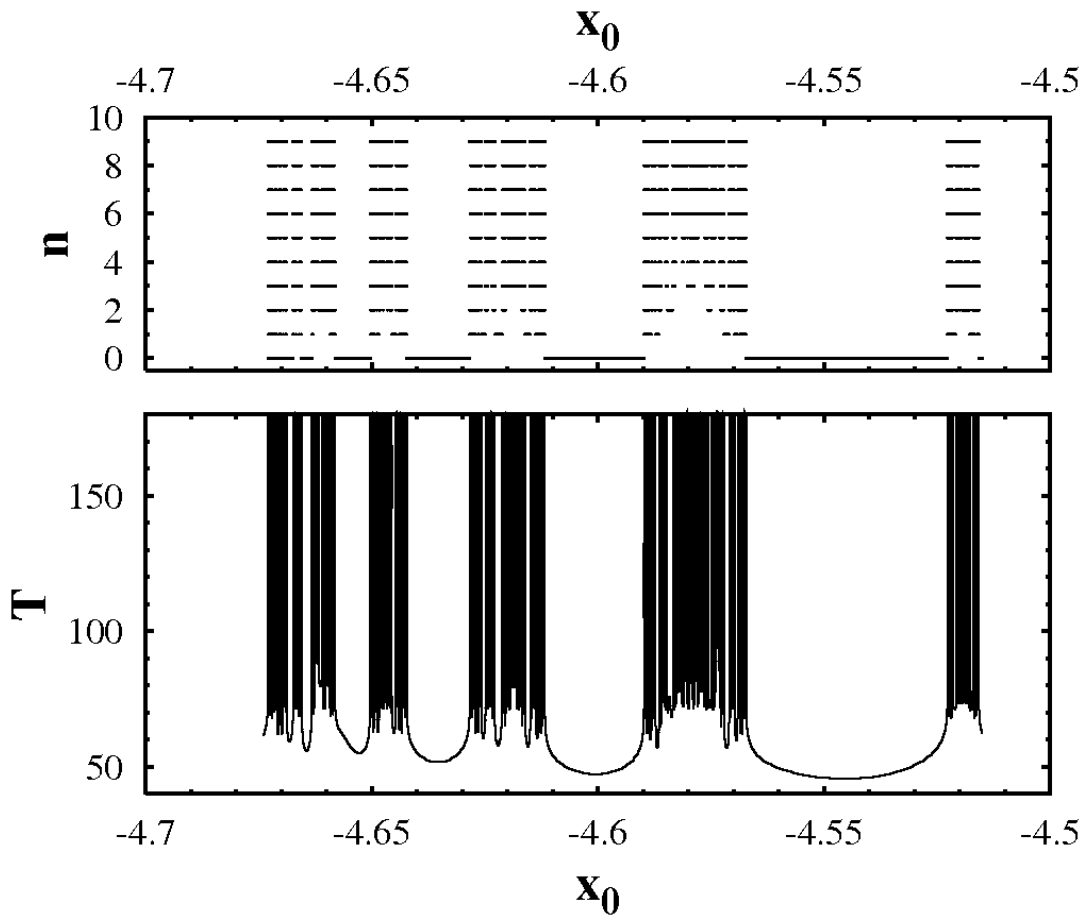}
\caption{Advection fractal under a high-frequency noisy perturbation.
(a) the number of turns of particles around the 
vortex $n$ and (b) the respective exit time $T$  vs initial particle's 
position $x_0$.}
\label{fig8}
\end{center}
\end{figure}
\begin{figure}[!htb]
\begin{center}
\includegraphics[width=0.8\textwidth,clip]{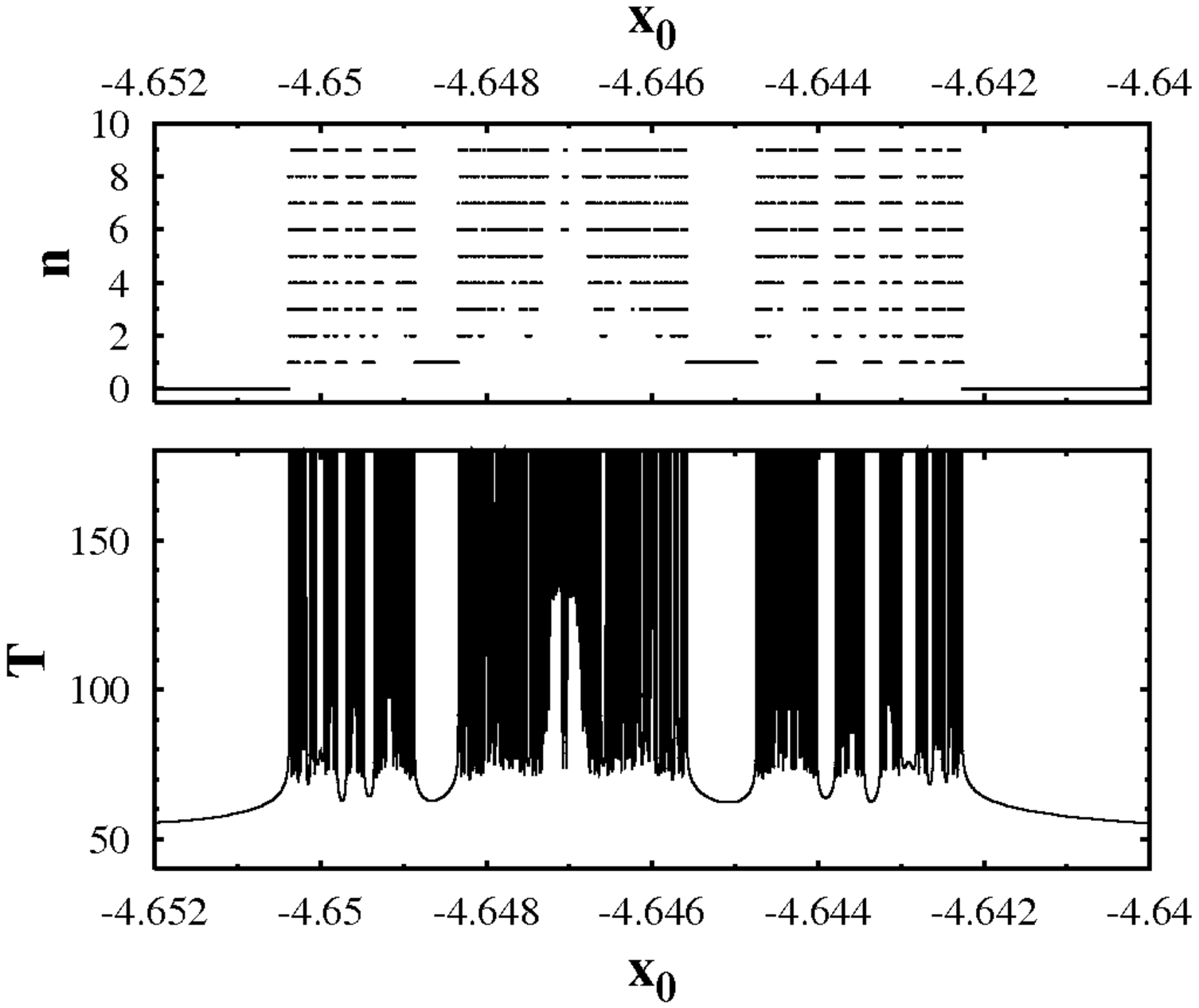}
\caption{Magnification of one of the epistrophes of the fractal in
Fig.~\ref{fig8}.}
\label{fig9}
\end{center}
\end{figure}
Let us consider transport and mixing of particles chosen on the material line 
in the incoming flow with the coordinates 
$y_0=-6$, $x_0\in[-4.674:-4.515]$ under the influence of a steady current 
with the velocity $\varepsilon =0.5$ and a noise with the amplitude 
$\xi=0.1$ and the frequency range $\omega\in[2:5]$. 
We compute 
the dependencies of the number of turns of particles around the 
vortex $n$ and the respective exit times $T$  on the initial particle's 
position. It follows from comparing Figs.~\ref{fig4} and~\ref{fig8},
that a hierarchical fractal
structure with epistrophes and strophes survives under a noisy 
excitation. Magnification of one of the epistrophes of the fractal in 
Fig.~\ref{fig8} 
demonstrates a self-similar structure (see Fig.~\ref{fig9}). 
Distribution of exit times $T$ under the noisy perturbation (not shown there) 
demonstrates a more heavy tail and increasing values of the exit times  
comparing with the case of purely periodic perturbation \cite{PD04}. 
Noisy-induced breaking of deep invariant curves in the vortex core 
occurs due to numerous resonances between frequencies of the noise and 
the unperturbed frequencies. Therefore, a large number of particles may 
diffuse into the vortex core and stay there for a long time.   
\begin{figure}[!htb]
\begin{center}
\includegraphics[width=0.8\textwidth,clip]{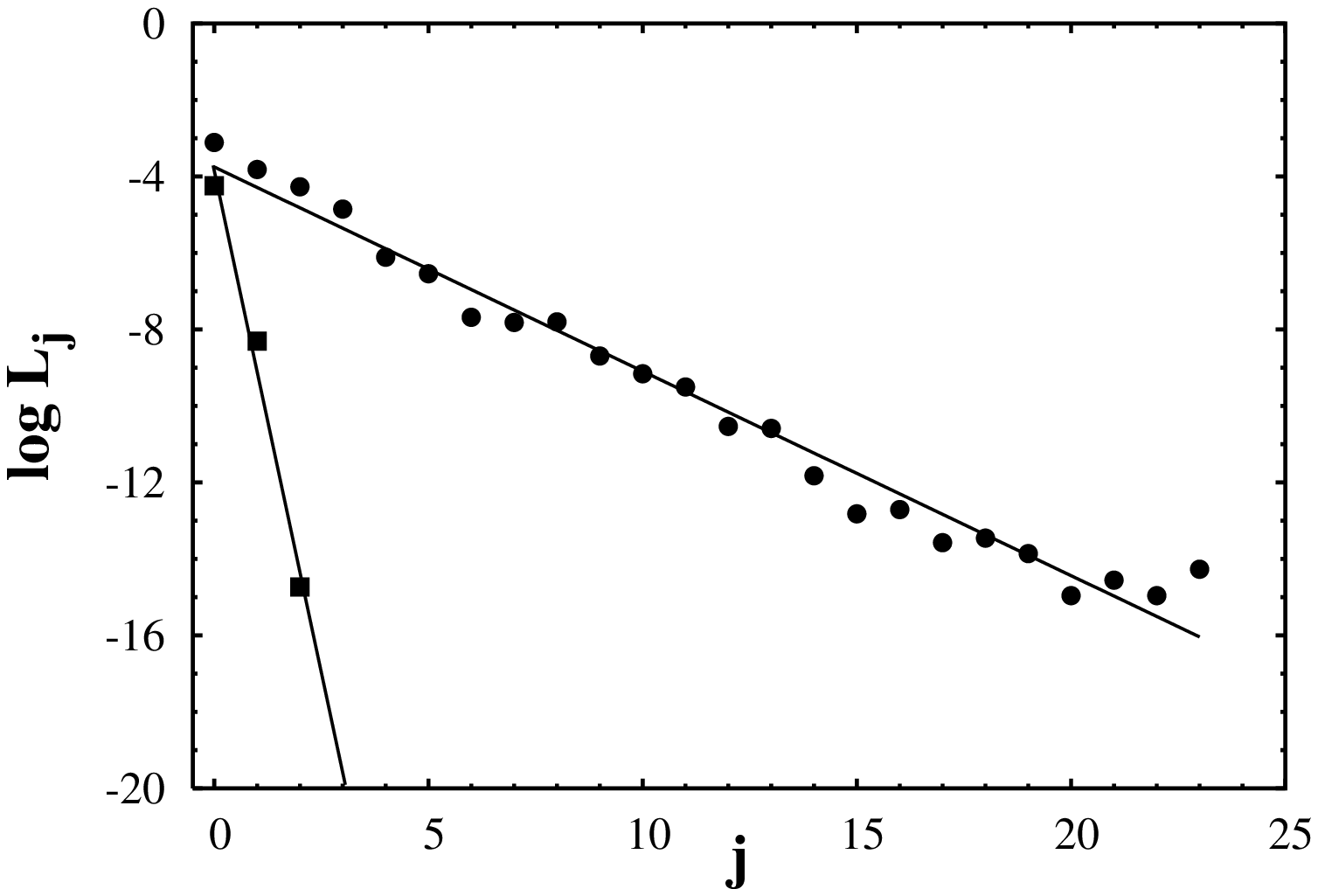}
\caption{Semilogarithmic plot of the decrease in the segment 
length $L_j$ on its number $j$ in the zero-level  epistrophe  
under  high-frequency (circles) and low-frequency (squares) 
noisy perturbations.}
\label{fig10}
\end{center}
\end{figure}
The epistrophic law, found in the preceding section with the flow with
a periodic perturbation, is valid under a noisy perturbation as well.  
In Fig.~\ref{fig10} we plot the dependencies 
of the segment lengths of the zero-level epistrophe on the number of the 
respective segment both for high-frequency ($\omega\in[2:5]$) and 
low-frequency ($\omega\in[0.4:1]$) noise. The exponential law is, in general, 
valid. We generalize the epistrophic law  in the case of noise as follows:
\begin{equation}
q=\exp(-2\pi\lambda/\omega_\text{eff}), 
\label{Epistrophic_law_ex}
\end{equation}
where $\omega_\text{eff}$ is an effective frequency of perturbation 
defining the average frequency of appearing lobe-like structures 
with which particles quit the mixing zone. 
Whereas this frequency is exactly equal to unity in a periodic flow   
(see Fig.~\ref{fig3}  with the lobe-like structures), it is equal to 
$\omega_\text{eff}\simeq 2.9$ under a high-frequency noisy perturbation 
and to $\omega_\text{eff}\simeq 0.3$ under a low-frequency one. 

\section{Noise-induced clustering of tracers}

In geophysical field experiments one works mainly with single 
realizations of stochastic processes of interest and on a finite time 
interval $[0:T_0]$. By these reasons we can recover ``deterministic'' 
methods to explore
the sets of stable trajectories satisfying to the condition of the
finite-time invariance:
{\it if any set in the phase space at $t=0$ transforms to itself
at $t=T_0$ without mixing, then
it corresponds to an ensemble of  
trajectories which are stable by
Lyapunov within the interval $[0:T_0]$}.
In order to find such stable sets for an arbitrary spectrum of perturbation,
we propose the following  map:
\begin{equation}
\displaystyle{x_{n+1}=x(x_n,y_n,T_0),\qquad
y_{n+1}=y(x_n,y_n,T_0)},
\label{ot}
\end{equation}
where $x_n$ and $y_n$~are initial particle's positions for the  $n$-th iteration.
In fact, this map is equivalent to a Poincar\'e map for a system under  
a periodic perturbation $\tilde F(\tau)$ consisting 
of identical pieces of $F(\tau)$ of the same duration $T_0$
\begin{equation}
\tilde F(\tau+mT_0)=F(\tau),\quad
0\le \tau\le T_0,\quad m=0,1, \dots .
\label{12}
\end{equation}
In this way we replace our original randomly-driven system by a
periodically-driven one. It should be emphasized that the validity 
of this replacement is restricted by the time interval $[0:T_0]$.
By analogy with the usual Poincar\'e map, the key property of the 
map (\ref{ot}) is the following:
{\it each point of a continuous closed trajectory of the 
map} (\ref{ot}) {\it corresponds to a starting point
of the solution of} Eqs.~(\ref{3}) 
{\it which remains stable by Lyapunov till the time $T_0$}.
The inverse statement is not, in general, true.
It will be shown below that the map (\ref{ot}) provides sufficient
but not necessary criterion of stability.

\begin{figure}[!htb]
\begin{center}
\includegraphics[width=0.71\textwidth]{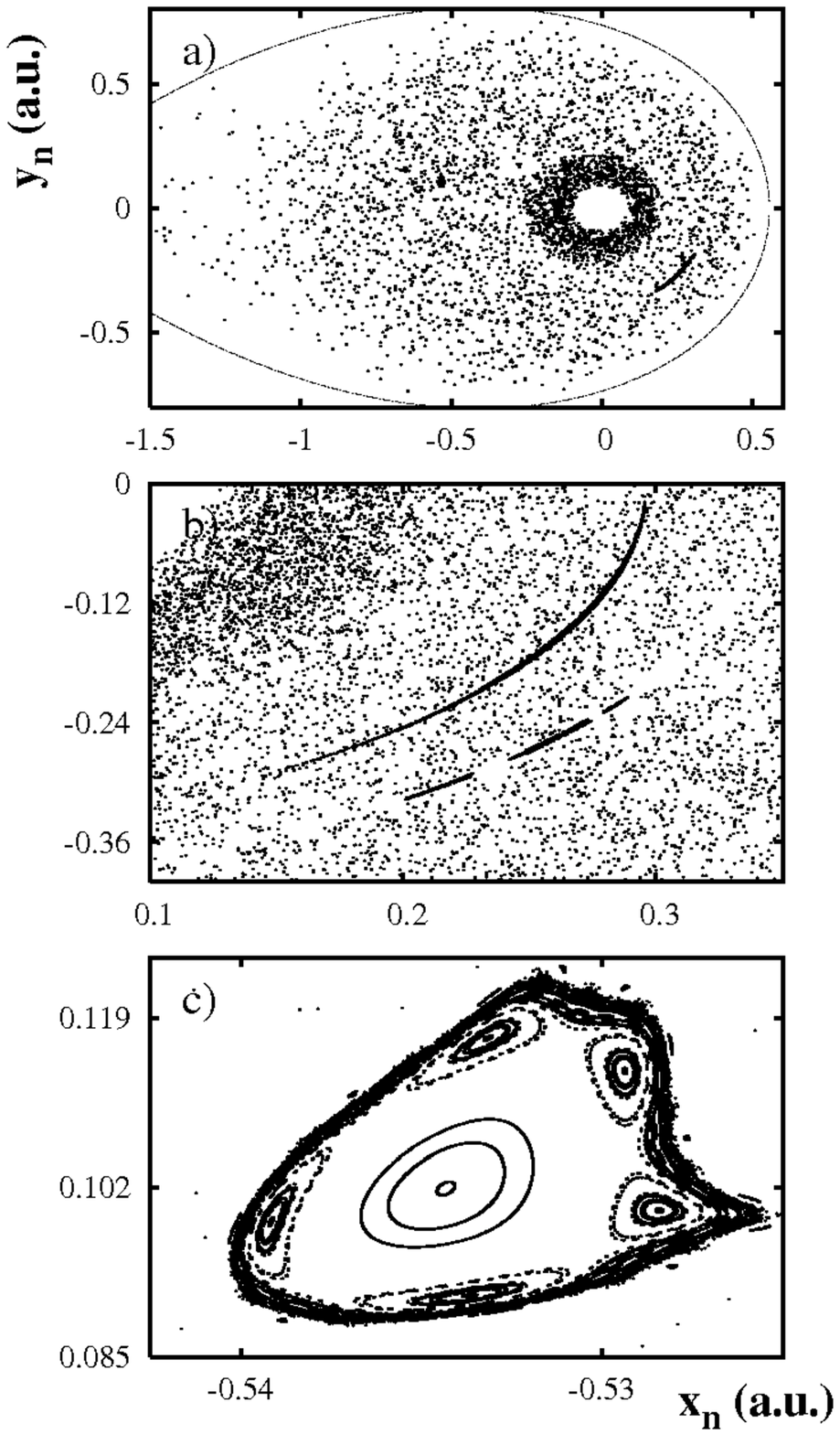}
\caption{The map (\ref{ot}) with $T_0=2.3\pi$ revealing clusters
of particles in a random velocity field with the amplitude
$\xi=0.1$ and the frequency band $\omega\in[9:10]$.
(a) general view in the phase plane $(x, y)$, (b) and (c)
magnification of the respective clusters in (a). The dotted line
is an unperturbed separatrix.
}
\label{fig11}
\end{center}
\end{figure}

\begin{figure}[!htb]
\begin{center}
\includegraphics[width=0.8\textwidth]{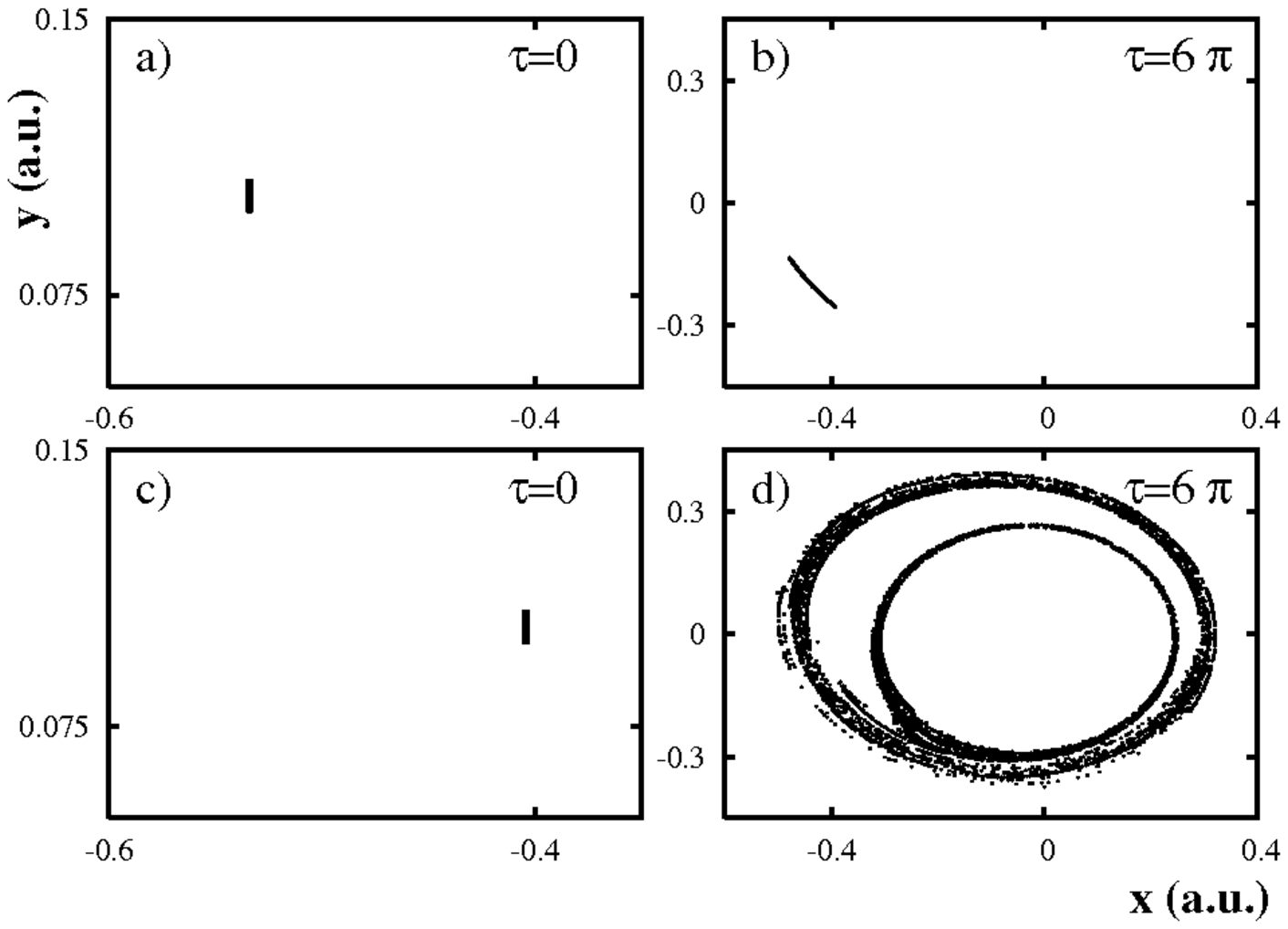}
\caption{The compact evolution of a coherent cluster of particles 
(the upper panel) and  
deformation of an unstable patch of particles (the lower panel) in a random 
velocity field. Parameters of noise are given
in caption to Fig.~\ref{fig11}.
}
\label{fig12}
\end{center}
\end{figure}

The phenomena, we report on in this section, is clustering
of passive particles in a noisy flow resembling clustering
of sound rays in an ocean underwater sound channel found in
\cite{MUP04}. We have developed a special method to seek for such clusters, 
i.~e. patches of tracers that move together for a long time in spite of
action of a broad-band noise. 
Fig.~\ref{fig11}a demonstrates 
the result of mapping with $T_0=2.3\pi$ on the configuration plane
$(x, y)$, and Figs.~\ref{fig11}b and c are magnification of the respective
clusters in Fig.~\ref{fig11}a. Prominent chains of the islands of stability 
can survive under a rather strong noisy perturbation ($\varepsilon=0.5$ and 
$\xi=0.1$) with the high-frequency range $\omega\in [9:10]$. 
To give a more direct manifestation of coherent clustering in a random flow 
that could be observed in real laboratory dye experiments, we 
compare the evolution of patches of particles  chosen in the  
regions of stability and instability in the configuration space. 
In the upper panel in  
Fig.~\ref{fig12} we show the evolution of a coherent cluster 
corresponding to the small black point (the region of stability for 
a given realization of noise) with coordinates 
$x \simeq -0.535$ and $y \simeq 0.1$ in Fig.~\ref{fig11}a. More or less  
compact evolution of this patch with $10^4$ particles goes on up to, 
at least, $\tau = 6\pi$. For comparison, the lower panel in Fig.~\ref{fig12}  
demonstrates on the same time interval the evolution of the patch 
with the same number of particles 
chosen initially close to the coherent cluster at 
$\tau =0$. The respective initial patch is deformed strongly at $\tau=6\pi$. 

We would like to emphasize that the duration of the temporal interval 
$T_0$ in constructing the map (\ref{ot})
can be chosen arbitrarily. Thus, if $F(\tau)$ is a stationary
random process, the regions of stability in the phase space exist at any time
moment. The map (\ref{ot}) enables to prove definitely the existence
of some regions of stability in the phase space but not all of them. 
Really, the map, 
constructed with any given value of $T_0$, can reveal only those stable sets
which correspond to the phase oscillations nearby the fixed points of the
map. However, there exist another regions of stability looking as chaotic
ones on the map. The topology of the map changes with varying the mapping
time $T_0$, and some regions in the phase space, which look as pseudochaotic on
the map at $T_0=T_1$, become stable at $T_0=T_2>T_1$. The total area of the 
regions of
stability, survived under a weak noise at $\tau=T_0$, can be estimated as 
an area of
superposition of all the stable sets detected by the map (\ref{ot}) with the
mapping step varying from $T_0$ to $T_c$, where $T_c$ is a mixing time.

Clustering in the ocean is a common feature that can be seen, for
example, in satellite images of the ocean surface. In resent
years new observation's tools~--- quasi-Lagrangian current following
floats and drifters~--- have been used to observe velocity
field in the ocean at different levels of depth (for a review see \cite{D91}).
In connection with our numerical observation of coherent clusters
of passive particles in the simple kinematic ocean model, we would
like to pay attention to the results of the SOFAR floats program in the
POLYMODE experiment in the North Atlantic \cite{RPW86}. Up to
forty neutrally buoyant floats at $700$~m and $1300$~m were used
to provide a quasi-Lagrangian description of the structure and evolution
of the mesoscale eddy field. Those floats can be considered as quasi-passive
tracers in a weak-noise environment. A large number of the deep floats revealed
remarkably coherent motion over a two-month period.

\section{Conclusion}
Kinematic models are attractive because of their simplicity, generality and
possibility to reveal fundamental geometric structures 
responsible for Lagrangian transport and mixing in the ocean. In this paper 
we have reviewed 
transport and mixing of passive particles in a simple kinematic two-dimensional
model with a topographical eddy and a tidal time-periodic current. We have  
found fractal features of transport and mixing in a kinematic model 
of the current-eddy interaction in a noisy environment and compared
the results obtained with the purely deterministic case.
The Lagrangian structures in our numerical
experiments have been shown to be similar to those found in a recent laboratory
dye experiment at Woods Hole \cite{DPH02}. We have demonstrated a 
non-trivial phenomenon of noise-induced clustering of passive 
particles and proposed a method to find the clusters in numerical experiments. 
These clusters are patches of advected particles which can move together in a
random velocity field for comparatively long time. The clusters appear due
to existence of regions of stability in the phase space which is the physical 
space in advection problems.

\section{Acknowledgments}
This work was supported by the Russia Government Program ``World Ocean'', 
by the Program ``Mathematical Methods
in  Nonlinear Dynamics'' of the Russian Academy of Sciences, 
and by the Program for
Basic Research of the Far Eastern Division of the Russian Academy of
Sciences.


\begin{thebibliography}{99}
\bibitem{D00} H.A. Dijkstra, Nonlinear physical oceanography,
Dordrecht, Kluwer, 2000.
%
\bibitem{B89} A.S. Bower, A simple kinematic mechanism for mixing fluid
parcels across a meandering jet, J. Phys. Oceanogr. 21 (1989) 173-180.
%
\bibitem{DW96} J.Q. Duan, S. Wiggins, Fluid exchange across a meandering jet
with quasi-periodic time variability, J. Phys. Oceanogr. 26 (1996) 1176-1188.
%
\bibitem{KK99} V.F. Kozlov, K.V. Koshel', Barotropic model of chaotic
advection in background flows,
Izv. AN. Fiz. Atmos. Okean. 35 (1999) 137-144 [Izvestiya, Atmospheric and
Oceanic Physics. 35 (1999) 123-130].
%
\bibitem{PD04} M. Budyansky, M. Uleysky, S. Prants, Hamiltonian fractals
and chaotic scattering by a topographical vortex and an alternating current,
Physica D. 195 (2004) 369-378. 
%
\bibitem{KJ04} L. Kuznetsov, C.K.R.T. Jones, M. Toner, Jr. A.D. Kirwan,
Assessing coherent-feature kinematics in ocean models,
Physica D. 191 (2004) 81-105.       
%
\bibitem{KK04} Yu.G. Izrailsky, V.F. Kozlov, K.V. Koshel, 
Some specific features of chaotization of the pulsating barotropic 
flow over elliptic and axisymmetric sea-mounts,
Phys. Fluids. 16 (2004) 3173-3190.
%
\bibitem{D91} R.E. Davis, Lagrangian ocean studies,
Annu. Rev. Fluid Mech. 23 (1991) 43-64. 
%
\bibitem{RPW86} T. Rossby, J. Price, D. Webb,
The spatial and temporal evolution of a cluster of SOFAR floats in the
POLYMODE local dynamics experiment (LDE), J. Phys. Oceanogr.
16 (1986) 428-442. 
%
\bibitem{JTPL01} M.V. Budyansky, S.V. Prants,
Universal mechanism of chaotic mixing in an elementary deterministic flow,
Pi{\'s}ma Zh. Tekh. Fiz. 27 (2001) 51-56
[Tech. Phys. Lett. 27 (2001) 508-510].
%
\bibitem{K83} V.F. Kozlov, Models of the Topographical Vortices in the
Ocean, Nauka, Moscow, 1983 [in Russian].
%
\bibitem{Z95} V.N. Zyryanov, Topographical Vortices in Dynamics of Sea
Currents, IVP RAN, Moscow, 1995 [in Russian].
%
\bibitem{H73}  N.G. Hogg, On the stratified Taylor columns,
J. Fluid Mech. 58 (1973) 517-537.
%
\bibitem{GC97} D.R. Goldner, D.C. Chapman, Flow and particle motion induced
above a tall seamount by steady and tidal background currents,
Deep-Sea Research I. 44 (1997) 719-744.
%
\bibitem{JETP04} M.V. Budyansky, M.Yu. Uleysky, S.V. Prants,
Chaotic scattering, transport, and fractals in a simple hydrodynamic flow,
Zh. Eksp. Teor. Fiz. 126 (2004) 1167-1179 [JETP. 99 (2004) 1018-1027].
%
\bibitem{M03} K.A. Mitchell, J.P. Handley, B. Tighe, J.B. Delos
and S.K. Knudson, Geometry and topology of escape,
Chaos. 13 (2003) 880-891. 
%
\bibitem{DPH02} H.E. Deese, L.J. Pratt, K.R. Helfrich,
A laboratory model of exchange and mixing between western boundary
layers and subbasin recirculation gyres,
J.~Phys. Oceanogr. 32 (2002) 1870-1889.
%
\bibitem{MUP04} D.V. Makarov, M.Yu. Uleysky, S.V. Prants,
Ray chaos and ray clustering in underwater acoustics
Chaos. 14 (2004) 79-95.
%
\end{thebibliography}
\end{document}